\documentstyle[aas2pp4,tighten,epsfig]{article}

\newcommand\etal{et al. }
\newcommand\degr{$^\circ$ }
\newcommand\Qsixty{Q$_{60}$}
\newcommand\per{$^{-1}$}
\newcommand{\ha}{H$\alpha$}
\newcommand{\hi}{H{\sc i}}

\newcommand{\htwo}{H$_2$}

\begin{document}
\baselineskip=15pt        

\title{Spatial distribution of far infrared emission in spiral galaxies
I. Relation with radio continuum emission}
\author{Y.D.\,Mayya{$^{1,2}$} and T.N.\,Rengarajan$^2$}
\affil{$^1$Tata Institute of Fundamental Research, Homi Bhabha Road, 
        Mumbai 400 005, India}  
\affil{$^2$Instituto Nacional de Astrofisica Optica y Electronica,  
           Apdo Postal 51 y 216, 72000 Puebla, Pue., M\'EXICO   }   
\affil{Electronic Mail: ydm@inaoep.mx  and renga@tifrvax.tifr.res.in}
\authoremail{ydm@inaoep.mx  and renga@tifrvax.tifr.res.in}
\vskip 0.5cm
\affil{\it Accepted --- June 1997. 
To appear in Astronomical Journal, September 1997}


\begin{abstract}

We use high resolution IRAS and 20~cm radio continuum (RC) images of 
a sample of 22 spiral galaxies to study the correlation between the
far infra-red (FIR) and RC emissions within the galactic disks.
A combination of exponential and gaussian profiles rather than a single
exponential profile is found to be a better representation of the 
observed intensity profiles in the two bands. The gaussian component, 
which we show is not due to the effects of limited beam-resolution, 
contains more than 60\% of the total flux in majority of the galaxies. 
The dominance of the gaussian component suggests that the nuclear star
forming regions and the bulge stars are more important contributors to the
emission in the two bands, rather than the outer exponential stellar disks.
The RC profile is flatter compared to the FIR profile, resulting in a 
decrease of their ratio, \Qsixty, away from the center. However, 
the \Qsixty\ increases in the extreme outer parts, where the dispersion 
in the FIR and RC correlation is also higher than in the central regions.
The global \Qsixty\ and its dispersion match those in the inner
parts of the galaxies. These results imply that the observed tight correlation 
in the global quantities reflects processes in the inner regions only
where OB stars and the associated Type~II supernovae control the FIR and 
RC emission. In the outer parts heating of very small dust grains by 
the old disk stars provides a secondary component in the FIR emission, without
associated RC emission.

The edge-on galaxy NGC\,3079 shows extended FIR and RC emissions along 
its minor axis, probably associated with the nuclear starburst activity. 


\end{abstract}

\section {Introduction}    

In recent years considerable efforts have been put in to improving the 
resolution of images obtained with the Infrared Astronomical Satellite (IRAS) 
by using different deconvolution techniques (Aumann et al. 1990; 
Ghosh et al. 1993). The resultant resolutions enable the study of the 
physical properties within galaxies over kpc scales in nearby galaxies.
Devereux and collaborators have made use of these high resolution images to
establish a good spatial correspondence
between the deconvolved IRAS images of the nearest galaxies 
and indicators of massive star formation as traced by \ha\ emission line
(M\,31: Devereux et al. 1994; M\,101: Devereux \& Scowen 1994, 
M\,81: Devereux et al. 1995). 
At the presently available resolution, the distribution of various 
physical quantities related to the far infrared emission can be studied
within galaxies at least up to the distance of the Virgo Cluster. 
We here undertake such a study and intend to find answers to the 
following questions: \\
1. Does the correlation seen between the global far infrared (FIR) and radio 
continuum (RC) emissions extend to local scales within galaxies?\\
2. How is the warm dust temperature distributed within galaxies? 
How significant are the contributions of the bulge and disk stars in heating
the dust? \\
3. How is the dust mass distributed within galaxies? How is it related 
to the gas mass? \\
The above questions become particularly important considering the
present understanding in this field based on global fluxes, which are
summarized below.\\ 
1. The observed correlation between the FIR and RC fluxes can
be understood in terms of phenomena associated with massive 
star formation, but the tightness of the correlation remains a puzzle.\\ 
2. While a majority of investigators believe the heating of 
dust by ultraviolet photons from young stars as a predominant 
source of the FIR emission (Helou et al. 1985; 
Sanders et al. 1986; Rengarajan \& Verma 1986; Devereux  et al. 1994), 
there have also been discussions on the role of heating by 
non-ionizing photons from the bulge and disk stars 
(Walterbos \& Schwering 1987; Smith et al. 1994; Xu \& Helou 1996).\\
3. Global data reveal gas-to-dust mass ratios in galaxies are around
5--6 times the value in the local Galactic surroundings 
(Devereux \& Young 1990). 

We have undertaken a comprehensive study of spiral galaxies based on the
available published data of the distribution of gas and RC emission 
in addition to IRAS data at $\sim$1\arcmin\ resolution. The first of the 
questions above is addressed in the present paper, while the other two 
questions are addressed in a companion paper (Mayya \& Rengarajan 1997; 
Paper II henceforth). The sample and analysis method adopted are discussed 
in sec.~2. In sec.~3, we present and discuss the 2-dimensional maps of the 
RC and 60\micron\ emission along with the fits to the azimuthally averaged 
radial intensity profiles. The variations in the 60\micron\ to RC flux ratio 
within the galaxies are also presented in this section. Issues related to 
the FIR-RC correlation are discussed in Sec.~4. One of the program 
galaxies (NGC\,3079) shows a possible FIR halo, which is discussed in sec.~5.
The main results of this study are summarized in sec.~6. Preliminary
results based on a smaller sample have been presented by us earlier
(Mayya \& Rengarajan 1996). 

\section{The sample, data and analysis procedure}   

We aim to study the distribution of the FIR
emission within disk galaxies and its relationship with the RC emission,
heating sources and the distribution of gas mass. Accordingly a list of
spiral galaxies is drawn for which the following data are
already available in a form convenient to use.\\
1. 20~cm RC images from the Very Large Array (VLA) at 1\arcmin\ resolution 
 (D configuration). \\
2. Radial surface density profiles of \hi\ and \htwo\  
  at a spatial resolution of $\sim$1\arcmin, either in tabular form or as 
  radial plots. \\
3. Optical angular diameter $>5$\,\arcmin. \\
4. 60\micron\ flux density  $>5$~Jy. \\

The last two criteria together ensure that there are a significant number 
of resolution elements in each image, and that the galaxies are bright 
enough in the IRAS bands to get a reliable High Resolution (HiRes) image. 
These selection criteria resulted in a sample of 22 spiral galaxies with 
a mean distance of 11~Mpc. The sample is not complete, but is representative 
of galaxies with a mean resolution of 3~kpc. The nearest of the 
galaxies such as M\,31 and M\,33 offer ten times better spatial
resolution but we did not consider using them in order to have a uniform 
spatial resolution in the sample galaxies. The basic properties of the 
sample galaxies are given in Table~1. The galaxy type, $B$ band magnitude, 
the optical diameter at 25~mag\,arcsec$^{-2}$ in arcmins, the axis 
ratio {\it b/a} and the position angle of the major axis have been taken 
from the {\it Third reference catalogue of bright galaxies} 
(de Vaucouleurs \etal\ 1991; RC3 henceforth). Distances are taken from 
the Nearby Galaxy Catalog (Tully 1988), wherein a Hubble constant 
of 75~km\,s$^{-1}$\,Mpc$^{-1}$ is assumed. The 60\micron\ beam size 
expressed in arcsec is the final resolution achieved after HiRes processing. 
S$_{60}$ and \Qsixty\ denote the total 60\micron\ flux density and the ratio
of total 60~\micron\ to 20~cm fluxes respectively. The table also 
contains FIR and RC diameters for individual galaxies, which will be 
discussed later.

We acquired the HiRes images (Rice 1993) of the program galaxies 
in four IRAS bands from the Infrared Processing and Analysis Center. 
HiRes uses Maximum Correlation technique (Aumann et al. 1990), which is
designed to recover the spatial information contained in the overlapping
detector data samples of the IRAS all-sky survey scans. The final
resolution is around 1\arcmin\ at 60 and 100\micron\ wavelengths. 
Resolution is better at 12 and 25\micron\ bands, but the sensitivities in 
these bands are too poor to trace out the outer regions in most of the
galaxies in our sample and hence these images have only limited use for our 
study. The HiRes processing was done using
the default configuration, which results in frames of 
1\degr$\times$1\degr\ field with 15\arcsec\ pixels. Sub-images of
60$\times$60 pixels are extracted for all galaxies except the nearest galaxy 
in the sample, NGC\,2403, for which sections of size 100$\times$100 pixels
are extracted. In all the cases images from the 20th iteration are used.
The VLA D configuration radio continuum data at 1.49~GHz (20 cm), are
collected for the program galaxies from the VLA data base (see Condon 1987; 
Condon et al. 1990). 
These RC images have gaussian beams of FWHM of either 48\arcsec,
54\arcsec\ or 60\arcsec\ for the sample galaxies and a pixel size of
14\arcsec. The fields identical to the FIR images are extracted, matching the
pixel sizes to 15\arcsec. The central coordinates in the extracted fields
correspond to the optical centers of galaxies as given in RC3.
Flux scale of RC images are converted from Jy\,beam\per\ to $10^6$Jy\,sr\per\ 
(or MJy\,sr\per), to match the flux units of FIR images. 

Surface brightness profiles of the HiRes and RC images are obtained by 
azimuthally averaging over elliptical annuli of 1\arcmin\ width. The 
ellipticity and position angle of the annuli are chosen based on the optical 
data in Table~1. Alternative analysis is also done in which every
pixel of 1\arcmin\ size, which is the typical resolution element, is treated
as an independent region for computing the desired quantities.
In this case, all the images are smoothed with a boxcar filter of width 
5$\times$5 pixel. The smoothed images are
rebinned into 1\arcmin\ pixels before obtaining the ratios of the images.
Ratios are neglected from further analysis, if the pixel values in any one of 
the bands is below 3 times the root mean squares (rms) noise in the image 
in order to avoid ratios involving small numbers.

The routines under Image Reduction and Analysis Facility (IRAF) and Space 
Telescope Science Data Analysis System (STSDAS) are used in the reduction and 
the analysis of the data. 

\section{Spatial distribution of FIR and RC emission}   

In one of the early comparisons of the FIR and the RC 2-dimensional structures
in galaxies, Beck \& Golla (1988) found a good correlation between the RC and 
FIR emission within the galaxies M\,31, M\,33, M\,101 and IC\,342. The
positional correspondence between the images in the two bands was also
found by Bicay et al. (1989) in spiral galaxies NGC\,5236 and 6946. 
The 1\arcmin\ resolution of the HiRes images allows such studies to be 
carried out on many more spiral galaxies. Marsh \& Helou (1995) 
extracted azimuthally averaged radial profiles of the HiRes and RC images  
and confirmed the larger scale lengths of RC intensity 
profiles found in the earlier low resolution studies (Bicay \& Helou 1990). 
They fitted the RC and FIR intensity profiles with exponential
functions, invoking a central unresolved source in some cases. 
In this section, we compare the structural correspondences between the FIR 
and RC images, in addition to analysing the radial intensity profiles. 

\subsection{Two-dimensional structures}   

The 60\micron\ and 20~cm surface brightness contour maps of the program 
galaxies are reproduced in Fig.~1. The images are aligned with north at
the top and east on the left. The tick marks on the plot boundary 
are in pixel units (15\arcsec). The horizontal bar at the bottom-left 
corner corresponds to a spatial scale of 5~kpc at the assumed distances 
to the galaxies. 
The highest and lowest contour values (in MJy\,sr\per) are marked on 
the bar chart on the right hand side of each panel. 
Starting with the peak value as maximum, consecutive contours are scaled 
down by factors of two until reaching a value which is around one sigma
above the background. This resulted in 7--10 contour levels in each plot. 
The optical major and minor axes are superimposed on the contour maps.
The HiRes and RC beams at ${1\over2}$ and ${1\over8}$ of the peak value are
shown as two concentric ellipses inside a square box on the top-left corner 
of each panel. Note that the beam contours are denoted by thick lines, so
as to distinguish them from the image contours when there is
overlap. The beam ellipses are smaller than the
corresponding image ellipses implying that the FIR and RC emitting
sources are resolved. The effect of resolution on the
extracted profiles is discussed in detail later in this section.
Compact background sources when present are easily 
distinguishable on RC images, and hence no attempt is made in removing these 
sources from the images. 

The striking resemblence between the 60\micron\ and 20~cm maps is easily 
noticeable in these maps. The major axis of the distribution in these two
bands is well aligned with the optical major axis, in general.
The FIR and RC derived axis ratios generally agree with the optical 
measurements, with one notable exception --- NGC\,3079 has a larger extension 
along the minor axis in both the FIR and RC compared to its optical value. 
The fact that the HiRes beam for this galaxy was among the best 
adds significance to the above result. This is a nearly edge-on galaxy 
and is known to show a variety of off-planar structures and hence a 
more detailed investigation of the observed FIR extension is 
done separately in Sec.~5.
The FIR and RC sizes of the galaxies are measured at the 0.60~MJy\,sr\per\  
and 0.01~MJy\,sr\per\ ($\sim$1 mJy\,arcmin$^{-2}$) levels on
the 60\micron\ and 20~cm surface brightness profiles respectively. 
These levels correspond to 3--6 and 2--5 times the rms noise 
above the local background on the FIR and RC images respectively. 
The resulting diameters are given in Table~1 as D$_{\rm{fir}}$ and 
D$_{\rm{rad}}$. The sizes at these levels generally match 
the optical sizes at 25 B~mag\,arcsec$^{-2}$ (D$_{\rm{opt}}$). However it
should be remembered that the measured sizes are overestimates due to the
smearing introduced by the 1\arcmin\ beam. 

\subsection{Decomposition of the radial intensity profiles}   

The azimuthally averaged radial intensity profiles at 60\micron\ and 20~cm are 
fitted with a function of the form,

\begin{equation}
I(r) = I_{\rm gau} exp(-(r/\sigma_{\rm g})^2) 
     + I_{\rm exp} exp(-(r/\sigma_{\rm e})  ),
\end{equation}
where the two terms represent gaussian and exponential components respectively.
The values of the four coefficients in the above equation are evaluated 
independently for the 60\micron\ and 20~cm profiles by using a least squares
fitting method. The STSDAS task {\it nfit1d} is used for the fitting purposes.
Total flux contained in each component is evaluated for the best fit 
parameters. We denote the flux fraction in the gaussian and exponential
components by the symbols f$_{\rm gau}$ and f$_{\rm exp}$ respectively. 

The results of the fit for all the galaxies are shown adjacent 
to the contour maps in Fig.~1. The open circles at every 1\arcmin\  
interval denote the observed profile. Gaussian and exponential components 
and their sum are shown by the dashed, dotted and solid lines respectively.
The downward pointing arrows on the x-axis of the 60\micron\ plots denote 
the optical isophote radii of the galaxies. It can be seen that a good fit 
was possible throughout the optical extent of the galaxies in most cases. 
Each plot contains the fitted
values of $\sigma_{\rm g}$ and $\sigma_{\rm e}$ in addition to f$_{\rm gau}$. 
A missing $\sigma$ on the plot indicates that the flux fraction in the
corresponding component is less than 20\%. However, we point out that
there are cases where the fitted outer exponential component is reliable
even when it contains less than 5\% of total flux (e.g. NGC\,4736).
In Table~2, we give the values of $\sigma_{\rm g}$, 
$\sigma_{\rm e}$, and f$_{\rm gau}$ for both the FIR and RC profiles,
whenever the component seems to be distinct, irrespective of the flux
fraction contained in that component. An observational scale length 
$\sigma_{\rm obs}$ is obtained by measuring the width of the observed
profile at 1/e of the peak intensity. $\sigma_{\rm obs}$ is independent 
of the profile fitting and is expected to match the $\sigma_{\rm e}$ 
for a purely exponential profile. Values of $\sigma_{\rm obs}$ for the
RC profile and its ratio with the FIR scale length are given in the last
two columns. Before we discuss the significance 
of the two components we compare fits with two components with that 
obtained using only an exponential function.

\subsubsection{Fits with exponential component alone} 

Bicay \& Helou (1990) and more recently Marsh \& Helou (1995) assumed 
the intensity profiles at the FIR and RC bands to be exponential in form 
and obtained the scale lengths by fitting the exponential function to 
the observed data. They added an unresolved central source in a few 
cases where the galaxy is known to harbor a starburst or an active 
galactic nucleus. Only two of the program galaxies (NGC\,3079 and 3628) 
in our sample are known to contain central compact sources.
We fitted a single exponential function to the intensity profiles.
The resultant 60\micron\ scale lengths are compared with that tabulated by 
Marsh \& Helou (1995) for the eight galaxies in common, in Fig.~2. 
The agreement between the scale lengths is satisfactory. 
 
However fitting only an exponential component results in higher residuals. 
In other words the rms departures from the fit obtained by using 
only an exponential function for fitting are systematically
higher as compared with the 2-component fitting described above. 
This is illustrated in Fig.~3 where the histogram of the ratio of the 
residuals of exponential fits to the 2-component fits is plotted.
The ratio is greater than unity in all the galaxies, except
in NGC\,2403, 7331 (FIR fits) and 4321 (RC fit). Thus, the addition 
of a gaussian component results in a better fit, as compared to pure 
exponential fits in majority of the program galaxies.

\subsubsection{Beam effects on the profile decomposition} 

In the previous subsections, we demonstrated that the addition of a gaussian
component to an exponential component results in a better fit to the
observations, and that the gaussian is often the dominant component.
Exponentially decreasing intensity profiles are more common in galactic 
disks. Hence we check whether the observed gaussian shaped profiles 
can be produced by the beam effects on an intrinsically exponential profile.
We choose the best-fit single component exponential profile and convolve 
it with the beam profile of the galaxy. For the sake of simplicity,
convolutions are performed on 1-dimensional profiles along the major 
axis of the galaxies. Four galaxies with different degrees of gaussian 
domination are selected for representation in Fig.~4. NGC\,2903 and 5055 
represent well-resolved galaxies in the sample, where as NGC\,4254 and 4303 
are among the galaxies with the minimum number of resolution elements.
Beam-convolved exponential profiles are shown by the dotted lines;
circles represent the observed profile, and the beam is shown 
by a dot-dashed line. For comparison, the best-fit 2-component profile is
plotted as a solid line. The percentage of the total flux in the gaussian 
component and the ratio of the rms errors of the single-exponential to 
the 2-component fit are noted in each panel. It can be inferred from the
plots that the beam-convolved exponential profile is only marginally 
different from the exponential profile, which will appear as a straight 
line on the plot. In galaxies with a dominant gaussian component,
the observed profiles are closer to the gaussian rather than the convolved 
exponential, independent of the resolution. In galaxies 
with f$_{\rm gau}<0.5$, the gaussian component can be an artifact of 
poorer resolution, but in those galaxies with larger f$_{\rm gau}$, the 
gaussian component must be due to an intrinsically gaussian disk.

It is further noted that there are purely exponential-shaped profiles among
the poorly-resolved galaxies, and gaussian-dominant
profiles among the well-resolved galaxies. This indicates
that the profile shapes are independent of resolution.
From all these discussions we conclude that the gaussian component 
is definitely not due to an artifact of poorer spatial resolution,
whenever the component contains more than $\sim60$\% of the total flux.

\subsection{Implications of the two components} 

It can be seen in Table~2 that the gaussian component is significant
(i.e. f$_{\rm gau} > 0.6$ in either FIR or RC) in 17 of the 22 galaxies. 
The five galaxies in which the component is insignificant are,  
NGC\,3079, 3198, 4321, 4656 and 6946.
The gaussian component stands out distinctly more often in the RC than
in the FIR. There are 13 galaxies with more than 80\% of the flux in the
gaussian component in the RC compared to 5 in the FIR. The mean flux 
fraction in the gaussian component is $0.62\pm0.25$ in the FIR, where as
it is $0.70\pm0.33$ in the RC. The mean of the difference 
f$_{\rm gau}$(RC)$-$f$_{\rm gau}$(FIR)$=0.08\pm0.30$. 
The reason for ${\rm f}_{\rm gau}$ being greater than ${\rm f}_{\rm exp}$ 
on an average may be that it is fitted as a 
central component in a profile which increases steeply towards the center. 

We now investigate the physical origin of the gaussian component. 
Firstly, we examine whether this component is due to an
unresolved central source. The ratio of full width at half maximum
(FWHM) of the gaussian component to the point source FWHM is plotted in 
Fig.~5. 
The gaussian component is clearly resolved both in the FIR and RC; 
thus a central point source cannot be responsible for the gaussian component. 
Hence we examine a possible relationship between the basic galaxy
parameters and the strength of the gaussian component. 
The flux fraction in the gaussian component is plotted 
as a function of Hubble type and inclination in Fig.~6 for the RC and
FIR separately. Barred, unbarred and intermediate galaxy types are 
distinguished by using different symbols. To within errors we see no 
effect of the bar or inclination. As a function of Hubble type also, 
there is no clear dependence. 
However, the lower envelope of the spread seems to decrease with increasing 
Hubble type. The latest type spiral in our sample is NGC\,4656 (T=9) and the 
gaussian fraction for this is $<0.1$ in both the bands.
Hence the bulge, which is absent in late type galaxies, may have a role
to play in creating a gaussian component.

Smith et al. (1994) have established that the observed dust 
temperatures in the central regions of NGC\,4736 and 3627 cannot be 
reproduced by the inferred massive stars there or by the nuclear activity, 
where as bulge stars can easily do the heating because of their large number.
Our sample includes both of these galaxies and both show strong gaussian 
components in both the FIR and RC emission. Thus it is likely
that the bulge stars are primarily responsible for producing 
non-negligible FIR luminosity in the gaussian component. 
This implies that the gaussian strength should decrease with Hubble type,
in agreement with the trend seen in Fig.~6. There may be
additional sources present in some galaxies, leading to the scatter 
in the correlation between the gaussian strength and the bulge
strength. The most likely candidates for these additional sources are the
circumnuclear star forming regions, which have typical  sizes
of 1--2~kpc. The gaussian component may also be accentuated by the steep 
gradient of gas (and hence dust) towards the center in the FIR as observed
for several galaxies (Young \& Scoville 1991). In the radio continuum
the increased magnetic field in the central region may contribute to the
dominance of the gaussian component.

The ratio of the observed scale lengths of the RC and FIR intensity 
profiles (last column in Table~2) is plotted 
against the fractional flux contained in the gaussian component in Fig.~7. 
The RC profiles are found to be broader than the FIR profiles in all
but four galaxies (NGC\,3628, 4192, 4321, 4569). The mean value of the 
ratio $\sigma_{\rm obs}$(RC)/$\sigma_{\rm obs}$(FIR) $=1.22\pm0.22$. 
The FIR and RC profiles have mean scale lengths of 5.0~kpc and 5.9~kpc 
respectively. The four galaxies with lower RC scale length, as compared to
that of FIR, behave differently from the rest in the radial distribution 
of \Qsixty. This will be discussed in the next sub-section.

The higher scale length for the RC as compared to that for the FIR is 
consistent with the findings of Bicay \& Helou (1990) and Marsh \& Helou
(1995). The larger scale length of the RC is believed to be a
consequence of the diffusion of radio emitting electrons away from the
place of origin, viz. supernova remnants around the Type~II
supernovae from OB stars in the disc. In 6 galaxies
viz. NGC 2903, 3198, 3628, 4736, 5033 and 7331, we find that besides the
central gaussian component, an exponential component is also present
at radii greater than $\sim$5~kpc in both the FIR and RC profiles. It is 
mostly a minor component in terms of the contribution to the total flux, 
but has higher flux 
than the gaussian component at radii exceeding 5~kpc. In four of these
six galaxies the exponential scale length for FIR exceeds the RC scale
length. This implies that the FIR emission drops more slowly than the RC 
emission in the outer regions of the galaxies. 
This is also confirmed by the fact that in our sample of 22
galaxies, 20 have a FIR exponential component in addition to a central 
gaussian component whereas only 10 have this additional component in RC. 
This effect is also seen in the radial profile of \Qsixty, the ratio of 
emission at 60\micron\ to that in the RC, as discussed below.

\subsection{Variation of \Qsixty\ within galaxies}  

A ratio involving the fluxes in the FIR and the RC wavelengths  
is useful in assessing the tightness of the correlation between the two 
quantities within a given galaxy. Accordingly we define \Qsixty\ as the ratio 
of emission at 60\micron\ to that at 20 cm. We construct the 1-dimensional
profile as well as the pixel map of \Qsixty.

Azimuthally averaged radial profiles in 60\micron\ and 20~cm
surface brightness are used to obtain the radial profiles of \Qsixty.  
\Qsixty\ is found to decrease by only a factor of around 3 in a given galaxy, 
in spite of FIR and RC intensities falling by several orders of magnitude.
This result is consistent with the findings of Bicay \& Helou (1990). 
The variation of \Qsixty\ is understood in terms of 
the larger RC scale length, compared to the FIR scale length. 
\Qsixty\ values in individual galaxies are plotted in Fig.~8, both as a 
function of 60\micron\ surface brightness (I$_{60}$) and the radial 
distance from the center.
Note that I$_{60}$ increases towards the left, so that the \Qsixty\ vs 
I$_{60}$ plot can be directly compared with the 
radial gradients in \Qsixty\ shown on the adjacent plots. 
The open circles represent values over 1\arcmin\ pixels, where as
the thin solid line (\Qsixty\ vs I$_{60}$) and thick solid line connecting 
filled circles (\Qsixty\ vs R/R$_{25}$) show the azimuthally averaged profiles.
The \Qsixty\ values obtained from the global data are indicated by the 
dotted horizontal line. \Qsixty\ shows a tendency to decrease away from the center 
in all galaxies except in NGC\,4192, 4321 and 4569. In NGC\,3628, 
the \Qsixty\ peak is shifted by 1\arcmin\ away from the nucleus. 
Note that these are the four galaxies in which the FIR scale length is 
larger than the RC scale length. \Qsixty\ obtained from the global 
data, in general, correspond to the values in the central regions of galaxies. 
This is an indication to the fact that the global quantities are flux
weighted and hence represent the properties of high intensity central regions.
It is interesting to note that the galaxies NGC\,628 and 2403 
do not have a well defined central peak in the distribution of FIR and RC
intensity, and their global \Qsixty\ values match the mid-disk values
rather than the central values.

In 13 galaxies, \Qsixty\ reaches a minimum around R/R$_{25}\sim0.5$
and then either increase or remain flat. It is significant to note
that this behavior is shown by all galaxies which have an outer exponential 
FIR component without a corresponding RC counterpart. 
In 6 galaxies, viz. NGC\,4192, 4321, 4569, 4736, 5033 and 7331, the 
outermost \Qsixty\ value is larger than the central value. Results presented 
in the companion paper (Paper~II) help us to understand the detailed 
behavior of the \Qsixty\ profiles, which is the topic of the next section. 

\section{Discussion on the FIR-RC correlation} 

The tightness of the correlation between the RC and FIR intensities
depends on the sources responsible for their emissions 
and the life times of these sources determine the time scales for 
contribution to the emission. The closer the two time scales, the tighter is
the correlation (Rengarajan \& Iyengar 1990). 
The FIR emission originates from dust grains which are 
heated by the ultraviolet photons from young OB stars and
long-lived intermediate mass stars (Walterbos \& Greenawalt 1996).
It is known that the various features in the interstellar extinction 
curve can be explained only by a mixture of classical big grains (BGs), 
very small grains (VSGs) and long molecular chains such as polycyclic 
aromatic hydrocarbons (PAHs) (see D\'esert et al. 1990). VSGs absorb 
photons and get heated temporarily to about 100~K, independent of photon 
energy and flux. Hence the FIR emission is proportional to the recent 
star formation rate only if the emission from BGs dominates the total 
FIR emission. Applying the model of D\'esert et al.  (1990), we find that 
the contribution of the BGs to the total 60\micron\ emission systematically 
decreases away from the center; the contribution is less than 50\% at radii 
larger than half the disk radius (Paper~II). Thus the FIR emission is more 
closely tied to recent star formation rate (SFR) in the inner regions 
compared to the outer regions.

What about the RC emission? 
Assuming that the sources of radio continuum emitting cosmic ray
electrons are supernovae (SN), the progenitor stars of SN determine
the time scale of contribution. Type~II SN originate from massive OB
stars while Type~I SN result from less massive and long-lived stars
(Trimble 1985). Rengarajan \& Iyengar (1990) have argued that the tight 
correlation between the global FIR and RC emissions implies that 
the Type~II SN contribution dominates. This is consistent with the 
observation of van den Bergh et al. (1987) that the frequency of 
Type~I SN in spiral galaxies is only one fourth the frequency of Type~II SN.
Thus the RC emission is directly related to the massive SFR.

Both the FIR and RC emissions are controlled by the recent star formation 
only in the inner parts of the galaxies. In the outer parts, transiently 
heated VSGs provide a source of FIR emission independent of the local SFR.
This explains the lower gradient of the FIR profile
and the increase of \Qsixty\ in the outer regions of galaxies. 
Dispersion on the correlation is also expected to be higher in the outer parts.  
We demonstrate in Fig.~9 that this is indeed the case for our sample of 
galaxies. Local \Qsixty\ values at 25, 50, 75 and 100\% of the
disk radius are plotted in the top panel. Individual galaxies are 
denoted by open circles at each of the four radial positions, with their
mean values joined by solid lines. 
The rms values (in log units) over these mean values are noted 
on the plots. Values 1-rms above and below the mean are denoted by the
dotted lines. Galaxies with the largest deviations from the mean \Qsixty\
are identified by their NGC numbers. The dispersion is found to remain
constant at 0.20~dex inside half the disk radius; the dispersion 
increases to 0.49~dex at the outer edges of the optical disk.

Is the global correlation affected by the higher dispersion in the outer
parts of galaxies? In order to answer this question we computed the
\Qsixty\ values by dividing the FIR and RC fluxes over progressively
increasing radial zones. The results are plotted in the lower panel of Fig.~9.   
The dispersion of 0.20 dex in the global value corresponds to the 
dispersions in the inner half of the disk. We have noted earlier that the 
global \Qsixty\ values represent the values in the inner parts 
of the galaxies. Thus the tightness in the global correlation only
requires the correlation to be tight in individual bright regions. 

\section {FIR halo in NGC\,3079?} 

The FIR and RC emissions along the minor axis of  
NGC\,3079 extend beyond the optical disk. The stellar disk of this galaxy is
oriented nearly edge-on and hence the larger minor axis extension
implies a larger scale height (or a halo) of the FIR emitting dust,
and radio emitting electrons. RC halos have been noticed in several 
edge-on galaxies (Hummel \etal\ 1991), but we have seen no reports of FIR 
halos. There has been indirect evidence
for the presence of dust in the halos of galaxies based on the line of sight 
optical depth towards distant quasars (Zaritsky 1994). Since dust 
grains are formed in the disks of galaxies, one needs a physical
mechanism for their transport to the halo. Before we discuss the possible
ways by which dust can be carried and maintained in the halos of galaxies, 
we shall look into the dust temperature distribution along the major and minor
axes of this galaxy. The galaxy is a bright FIR source, and hence
the 25\micron\ image is also used in the analysis. We compare the minor 
axis flux profiles in 25, 60 and 100\micron\ bands with their beam profiles 
along the same direction in the top three panels of Fig.~10. 
These plots indicate that the observed minor axis structures are resolved.
The FIR flux ratios 
f$_{25}$/f$_{60}$ and f$_{60}$/f$_{100}$, which are indicators of 
dust temperatures, are plotted along the two axes of the galaxy in 
the bottom-most panel of Fig.~10. The observed points, which are 
averages over 1\arcmin\ regions, are joined by lines to illustrate the 
difference in the gradients along the two axes. 
The FIR flux ratios, 
especially the one involving 25\micron\ (the hotter dust), falls off faster 
along the minor axis. This is consistent with the picture that the 
heating sources are in the disk and hence the halo dust must be cooler
than that in the disk.

Non-thermal RC halos have been interpreted in terms of the escape of cosmic
ray electrons away from the disk of galaxies along the open magnetic field
lines or inflated flux tubes (see Parker 1992). In recent years, vertical 
dust structures extending up to several kpc, 
thought to be tracing the magnetic field lines, have been noticed in 
several galaxies (Sofue 1987). Using the computations of Franco et al. (1991), 
Sofue et al. (1994) have shown that the radiation pressure due to star
light from the disk and bulge, particularly from starburst regions, would
drive the dust grains (associated with partially ionized gas) along
vertical magnetic tubes producing vertical dust structures for magnetic
fields stronger than a few $\mu$G. Central regions of starburst galaxies
are preferred regions for the formation of vertical structures, 
due to their strong radiation pressure as well as the magnetic field strength.
The ejected gas and dust fall back into the disk at various radial zones,
thus enriching the halo with dust and gas on a global scale.

NGC\,3079 has been the target for the detection of extra-planar structures
in different bands. de Bruyn (1977) and Hummel et al. (1983) detected
radio continuum features normal to the plane of the galaxy. Irwin \etal\
(1987) and Duric \& Seaquist (1988) detected a wind associated with strong
magnetic fields. The galaxy is among a few
objects which show biconic ionized filaments centered around the nucleus --- a
manifestation of violent starburst activity in the galactic nucleus
(Heckman \etal\ 1990). 
NGC\,3079 is known to contain diffuse ionized gas (DIG) extending to
several kpc in the halo (Veilleux \etal\ 1995).  
Thus, NGC\,3079 satisfies all the conditions for the
presence of warm dust in its halo. However, it is interesting to note that,
NGC\,3628, another edge-on galaxy in the sample also shows evidence for
the presence of nuclear superwinds (Fabbiano \etal\ 1990), but does not
show any trace of FIR halo. 

\section{Conclusions}  

Azimuthally averaged radial intensity profiles in 60\micron\ and 20~cm 
continua bands are analyzed in a sample of 22 spiral galaxies. A combination 
of gaussian and exponential components is found to fit the observed radial 
intensity profiles in the two bands better than the conventional exponential 
component alone. The majority of the galaxies contain more than 60\% of 
the flux in the central gaussian component in both the bands. The sample 
galaxies show a predominantly gaussian component more often in the RC than
in the FIR. The RC scale lengths are larger than the FIR scale lengths, which
results in a radial decrease of \Qsixty, the flux ratio in the two bands. 
In a significant number of the galaxies, \Qsixty\ remains flat or increases 
beyond half the disk radius, which is caused by the outer exponential 
component in the FIR without a RC counterpart. The tightness of the global 
FIR-RC correlation is determined by the processes in the inner regions only.
All the observed properties can be understood by a model in which  
the kpc-scale nuclear star forming regions and the bulge
stars are responsible for the FIR and RC emission in the central parts. 
The very small grains heated by the old disk stars dominate the
FIR emission, without contributing to RC emission
due to the lack of Type\,II supernovae, in the outer galaxies. 

The edge-on galaxy NGC\,3079 shows FIR and RC halos, possibly associated 
with the nuclear starburst activity.

\begin{acknowledgements}

 It is a pleasure to thank Dr. Walter Rice, the referee, for his comments 
 and suggestions towards improvement of the manuscript.
 We thank Dr. J.J. Condon for supplying the CDROM containing RC data
 of galaxies, which formed the basis of this work. I (ydm) thank
 Dr. S.K. Ghosh for introducing me to the IRAS data-base and also to the
 various pipelines run by IPAC. TNR thanks Dr. G.G.Fazio and the hospitality
 of Smithsonian Astrophysical Observatory, Cambridge, USA where part of this
 work was done as a Short Term Visitor.
 This research has made use of the NASA/IPAC Extragalactic Database (NED)   
 which is operated by the Jet Propulsion Laboratory, California Institute   
 of Technology, under contract with the 
 National Aeronautics and Space Administration.                                                            
  
\end{acknowledgements}

\clearpage     


\begin{deluxetable}{llrrrrcrrrcr}
\tablenum{1}
\tablewidth{0pt}
\tablecaption{Basic data on sample galaxies}
\tablehead{
\colhead{NGC}           & \colhead{Type}      &  \colhead{B$_{\rm T}^0$}  &  
\colhead{dist}          & \colhead{S$_{60}$}  &  \colhead{Q$_{60}$}       &  
\colhead{$60\mu$m beam} & \multicolumn{3}{c}{Diameter ($^\prime$) }       &  
\colhead{b/a}           &  \colhead{PA ($^\circ$)} \\ 
\colhead{}           & \colhead{}      &  \colhead{}  &  
\colhead{Mpc}        & \colhead{Jy}    &  \colhead{}  &  
\colhead{$^{\prime\prime}$ x $ {^{\prime\prime}}$}    & 
\colhead{opt}        & \colhead{fir}   &  \colhead{rad}  &  
\colhead{opt}        &  \colhead{opt}  }  
\startdata
 628 & SA(s)c   &  9.76 &  9.7 & 25.5 &  142 & 77 x 42 & 10.47 & 9.0 & 10.0 & 0.91 &   25  \\
2403 & SAB(s)d  &  8.43 &  4.2 & 62.6 &  190 & 60 x 42 & 21.88 &15.5 & 12.5 & 0.56 &  127  \\
2841 & SA(r)b   &  9.58 & 12.0 &  6.3 &   75 & 79 x 45 &  8.13 & 7.3 &  8.5 & 0.44 &  147  \\
2903 & SAB(rs)bc&  9.11 &  6.3 & 67.6 &  166 & 68 x 45 & 12.59 &13.3 & 10.0 & 0.48 &   17  \\
3079 & SB(s)c   & 10.41 & 20.4 & 52.8 &   62 & 40 x 35 &  7.94 & 6.0 &  6.5 & 0.18 &  165 \\
3198 & SB(rs)c  & 10.21 & 10.8 & 6.9  &  250 & 67 x 32 &  8.51 & 8.0 &  7.0 & 0.39 &   35 \\
3627 & SAB(s)b  &  9.13 &  6.6 & 66.5 &  145 & 72 x 42 &  9.12 &12.0 &  9.5 & 0.46 &  173 \\
3628 & ScP      &  9.31 &  7.7 & 56.9 &  108 & 86 x 35 & 14.79 &12.5 & 14.0 & 0.20 &  104 \\
4192 & SAB(s)ab & 10.02 & 16.8 &  8.4 &  114 & 92 x 45 &  9.77 &11.5 & 10.0 & 0.28 &  155 \\
4254 & SA(s)c   & 10.10 & 16.8 & 40.7 &   96 & 64 x 43 &  5.37 & 6.8 &  7.0 & 0.87 &    0  \\
4303 & SAB(rs)bc& 10.12 & 15.2 & 40.2 &   97 & 83 x 45 &  6.46 & 6.3 &  7.5 & 0.89 &    0  \\
4321 & SAB(s)bc &  9.98 & 16.8 & 26.9 &   79 & 55 x 35 &  7.41 & 7.0 &  6.5 & 0.85 &   30 \\
4501 & SA(rs)b  &  9.86 & 16.8 & 20.7 &   75 & 61 x 44 &  6.92 & 6.5 &  7.0 & 0.54 &  140 \\
4535 & SAB(s)c  & 10.32 & 16.8 & 12.6 &  195 & 62 x 44 &  7.08 & 7.8 &  7.5 & 0.71 &    0  \\
4569 & SAB(rs)ab&  9.79 & 16.8 & 11.0 &  132 & 84 x 44 &  9.55 & 8.0 &  5.0 & 0.46 &   23 \\
4656 & SB(s)mP  & 10.10 &  7.2 &  7.0 &  112 & 76 x 44 & 15.14 &14.0 & 14.0 & 0.20 &   33 \\
4736 & RSA(r)ab &  8.75 &  4.3 & 75.2 &  296 & 63 x 42 & 11.22 & 5.8 &  5.5 & 0.81 &  105 \\
5033 & SA(s)c   & 10.21 & 18.7 & 21.6 &  121 & 65 x 43 & 10.72 & 8.5 &  7.5 & 0.47 &  170 \\
5055 & SA(rs)bc &  9.03 &  7.2 & 50.9 &  131 & 61 x 42 & 12.59 &10.0 &  9.5 & 0.58 &  105 \\
6503 & SA(s)cd  & 10.11 &  6.1 & 11.1 &  304 & 67 x 44 &  7.08 & 8.3 &  7.5 & 0.34 &  123 \\
6946 & SAB(rs)cd&  7.78 &  5.5 &165.2 &  118 & 39 x 35 & 11.48 &11.8 & 14.0 & 0.85 &   32 \\
7331 & SA(s)b   &  9.38 & 14.3 & 41.9 &  112 & 75 x 43 & 10.47 &14.0 & 11.0 & 0.36 &  171 \\
\enddata
\end{deluxetable}

\begin{deluxetable}{lrccrccrc}
\tablenum{2}
\tablewidth{0pt}
\tablecaption
{Results of fits to the radial profiles of galaxies\tablenotemark{1}}
\tablehead{
\colhead{NGC}           & \multicolumn{3}{c}{FIR} & \multicolumn{3}{c}{RC} &
\colhead{$\sigma_{\rm obs}$(RC)}  & 
\colhead{ ${\sigma_{\rm obs}{\rm (RC)}\over{\sigma_{\rm obs}{\rm (FIR)} }}$} 
\\
\colhead{}  &  
\colhead{$\sigma_{\rm g}$}  & \colhead{$\sigma_{\rm e}$}  & 
\colhead{f$_{\rm gau}$}     & \colhead{$\sigma_{\rm g}$}  & 
\colhead{$\sigma_{\rm e}$}  & \colhead{f$_{\rm gau}$}    &
\colhead{}  &  \colhead{}   } 
\startdata
628  &   6.5  & \nodata &0.90 &   8.2 &\nodata&  1.00 & 9.49 & 1.42 \\
2403 &   3.6  &  2.7 &   0.64 &   4.4 &\nodata&  0.88 & 4.39 & 1.29 \\
2841 &   7.6  & \nodata &1.00 &   8.9 &\nodata&  1.00 & 8.76 & 1.10 \\
2903 &   2.5  &  4.2 &   0.93 &   3.3 &   1.8 &  0.52 & 2.78 & 1.20 \\
3079 &  12.9  &  6.9 &   0.18 &  16.1 &   9.1 &  0.45 & 12.41& 1.41 \\ 
3198 &   4.9  &  4.6 &   0.60 &   6.2 &   4.4 &  0.28 & 5.56 & 1.17\\
3627 &   3.1  &  2.5 &   0.76 &   3.9 &\nodata&  1.00 & 4.00 & 1.34 \\
3628 &   4.6  &  2.6 &   0.74 &   5.2 &   3.0 &  0.24 & 3.72 & 0.91\\
4192 &  13.9  &  4.8 &   0.63 &  12.0 &   7.4 &  0.74 & 11.41& 0.98\\
4254 &   5.2  &  3.3 &   0.59 &   7.9 &\nodata&  0.88 & 6.87 & 1.48\\
4303 &   5.1  &  6.2 &   0.94 &   6.4 &\nodata&  1.00 & 6.30 & 1.23\\
4321 &   5.9  &  3.7 &   0.35 &\nodata&   3.1 &  0.00 & 4.33 & 0.85\\
4501 &   6.2  &  3.5 &   0.69 &   7.4 &\nodata&  0.98 & 7.47 & 1.31\\
4535 &   6.1  &  3.8 &   0.31 &   9.2 &\nodata&  0.84 & 6.14 & 1.20\\
4569 &   6.2  &  5.4 &   0.56 &   5.8 &\nodata&  0.99 & 5.75 & 0.90\\
4656 &\nodata &  5.4 &   0.05 &\nodata&   6.6 &  0.08 & 5.95 & 1.17\\
4736 &   1.0  &  3.4 &   0.98 &   1.5 &   1.4 &  0.95 & 1.57 & 1.88\\
5033 &   5.3  &  5.8 &   0.65 &   6.6 &   4.5 &  0.58 & 6.11 & 1.16\\
5055 &   3.1  &  2.1 &   0.44 &   4.9 &\nodata&  0.82 & 3.69 & 1.33\\
6503 &   3.2  &  1.6 &   0.66 &   3.6 &\nodata&  0.91 & 3.56 & 1.28\\
6946 &   1.1  &  1.7 &   0.39 &   0.9 &   2.3 &  0.20 & 1.50 & 1.14\\
7331 &   7.7  &  6.3 &   0.47 &   9.1 &   8.1 &  0.89 & 8.83 & 1.19\\
\tablenotetext{1}{$\sigma_{\rm g}$ and $\sigma_{\rm e}$ are in kpc.
$\sigma_{\rm obs}$(RC) and $\sigma_{\rm obs}$(FIR) are the widths 
of the observed profiles measured at 1/e of the peak intensity.
}
\enddata
\end{deluxetable}

\onecolumn


\begin{figure}[htb]
\caption{
60\micron\ and 20 cm emission contours and azimuthally averaged intensity
profiles of 22 galaxies are shown. The maximum and minimum contour levels
in units of MJy\,s\per\ are indicated on the bar chart at the right hand 
side of each contour plot. Starting with the peak intensity,
successive contours differ by a factor of two. The HiRes and RC beams 
at $1\over 2$ and $1\over 8$ of the peak level are shown as two 
concentric ellipses at the top-left corner of each panel. The horizontal 
bar at the bottom-left corner corresponds to a linear scale of 5~kpc
within the galaxies at the assumed distances. The optical major and
minor axes are shown by a cross centered on the galaxy image. 
The coordinate system of the contour plots is in pixel units (15\arcsec).
The two panels on the right hand side depict the observed intensity 
profiles at 60\micron\ and 20 cm, along with the two-component fits
described in the text. Circles represent the observed values sampled at every 
1\arcmin, where as the dashed and dotted lines represent gaussian and 
exponential component of the fit respectively. The solid lines represent 
the sum of the two components. 
The most important parameters obtained from the fit are indicated. 
The downward pointing arrow on the radius axis of 
60\micron\ radial profile denotes the optical radius for each galaxy.
}
\end{figure}

\begin{figure}[htb]
\vspace*{-2cm}
\centerline{\psfig{figure=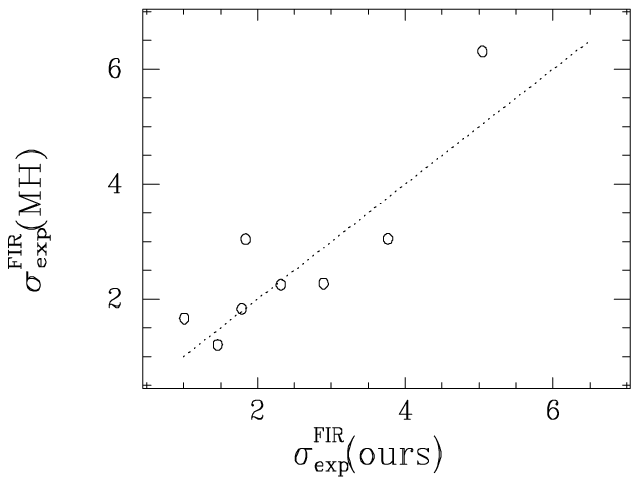,height=11cm}}
\vskip -1cm
\caption{
The exponential scale length (in kpc) for the FIR emission as tabulated by 
Marsh \& Helou (1995) is plotted against the value obtained by us for the
single exponential component for the 8 galaxies in common. 
The dotted line corresponds to a unity-slope line passing through the origin. 
}

\vskip -2cm
\centerline{\psfig{figure=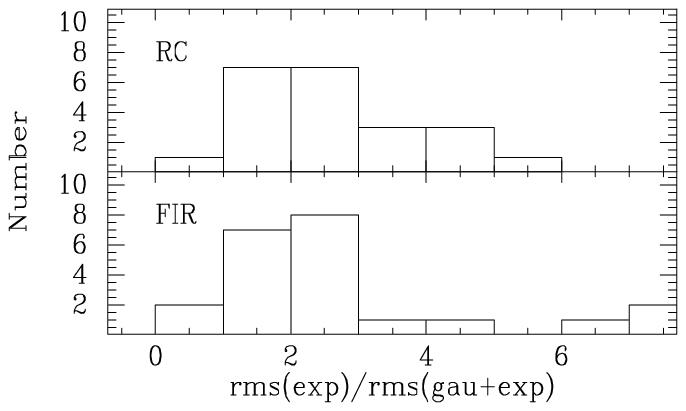,height=11cm}}
\vskip -2cm
\caption{
The root mean squares (rms) residuals from the 2-component fits, shown in
Fig.~1, are compared with the rms residuals obtained by one-component 
(exponential) fits. Except for two galaxies in FIR (NGC\,2403, 7331) and 
one in RC (NGC\,4321), 
the ratio is greater than unity, implying the necessity of a gaussian 
component in addition to the exponential component to represent the emission 
profiles in the FIR and RC bands.
}
\end{figure}

\begin{figure}[htb]
\vspace*{-2.0cm}
\centerline{\psfig{figure=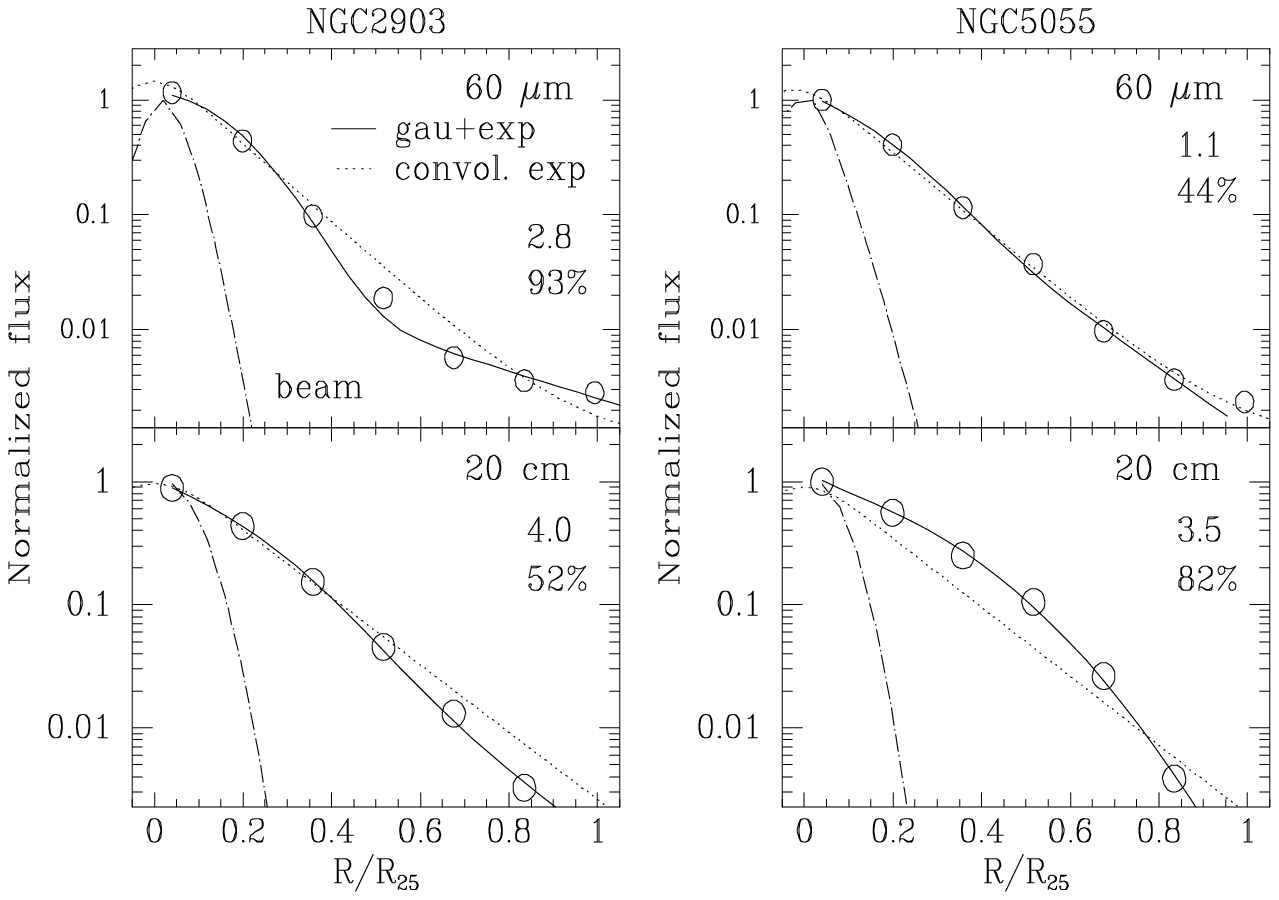,height=11.0cm}}
\vspace*{-1.5cm}
\centerline{\psfig{figure=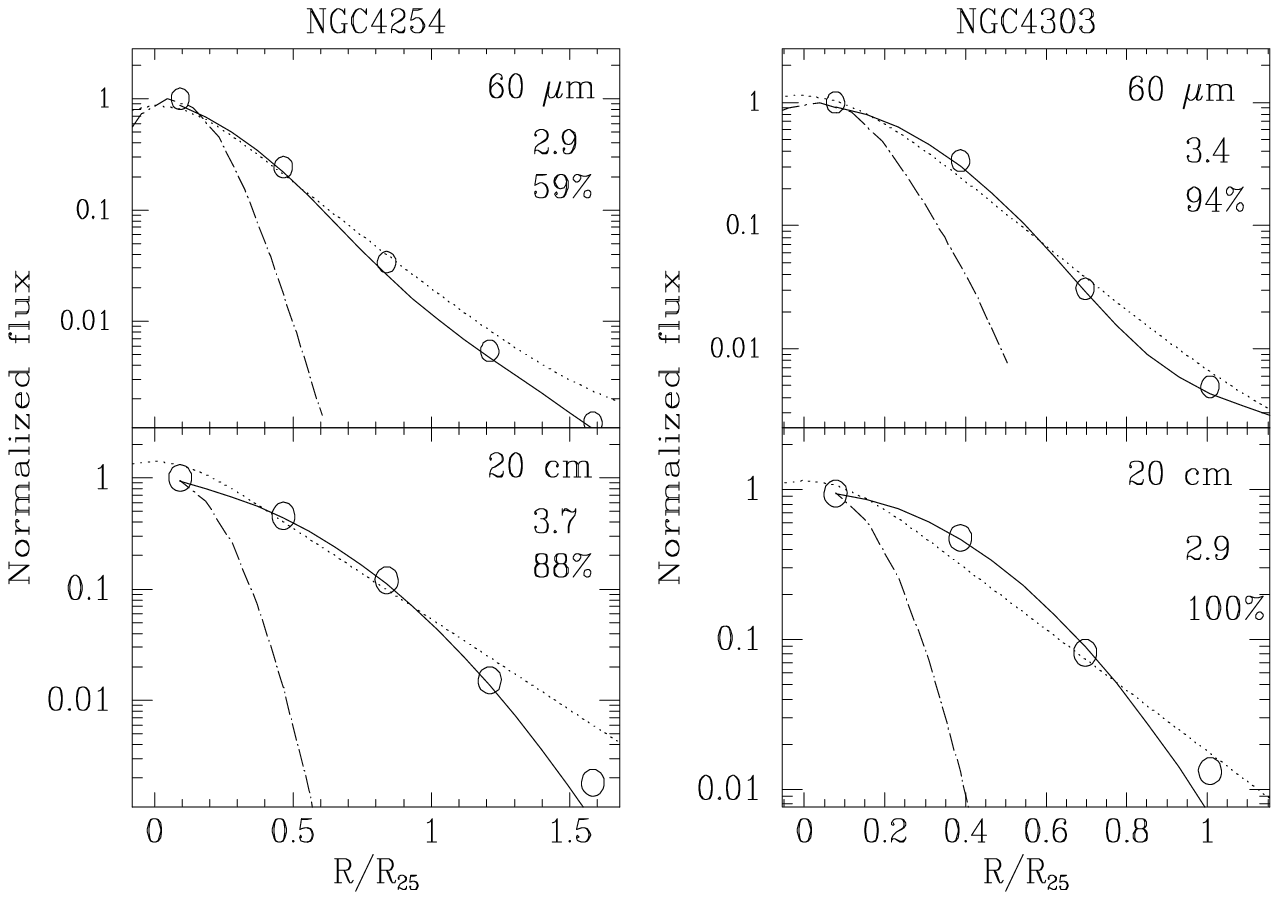,height=11.0cm}}
\vspace*{-0.5cm}
\caption{
The beam-convolved exponential profile (dotted line) is compared with the
best-fit 2-component profile (solid line) and observations (circles).
The beams used for convolution are shown by dot-dashed lines. 
The ratio of the rms errors of the single exponential component to 
that of the 2-component and the percentage flux in the gaussian component 
are indicated on the plot. 
NGC\,2903 and 5055 represent well-resolved galaxies, where as NGC\,4254
and 4303 represent marginally resolved galaxies. See text for details.
}
\end{figure}


\begin{figure}[htb]
\vspace*{-2cm}
\centerline{\psfig{figure=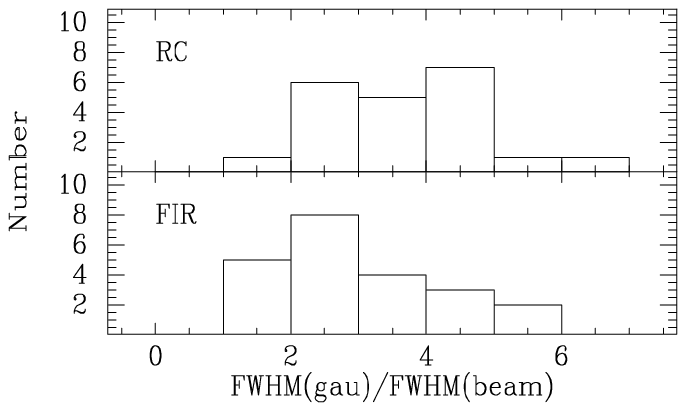}}
\vspace*{-2cm}
\caption{
The full width at half maximum of the gaussian component is compared with
that of the resolution beam. All the galaxies have the ratio greater
than unity, implying that the gaussian component is not due to a
compact unresolved source. 
}

\centerline{\psfig{figure=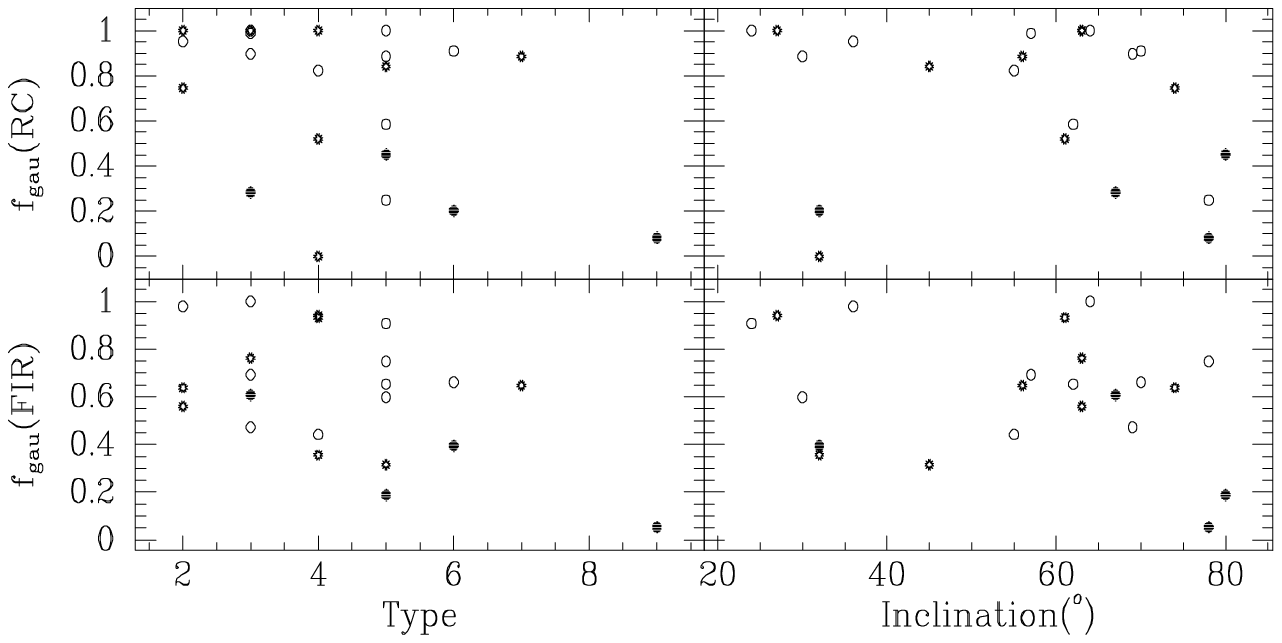}}
\vspace*{-2cm}
\caption{
Flux fractions in the gaussian component (f$_{\rm gau}$) for the FIR and RC
emissions are plotted against the Hubble type and inclination of the
galaxies. Barred, un-barred and intermediate type galaxies are distinguished
by the filled, open and starred circles respectively. Though a clear trend
is missing in any of these plots, the lower envelope seems to decrease as
Hubble type increases. The late-type galaxy 
NGC\,4656 has the least f$_{\rm gau}$.
}
\end{figure}

\begin{figure}[ht]
\vspace*{-3cm}
\centerline{\psfig{figure=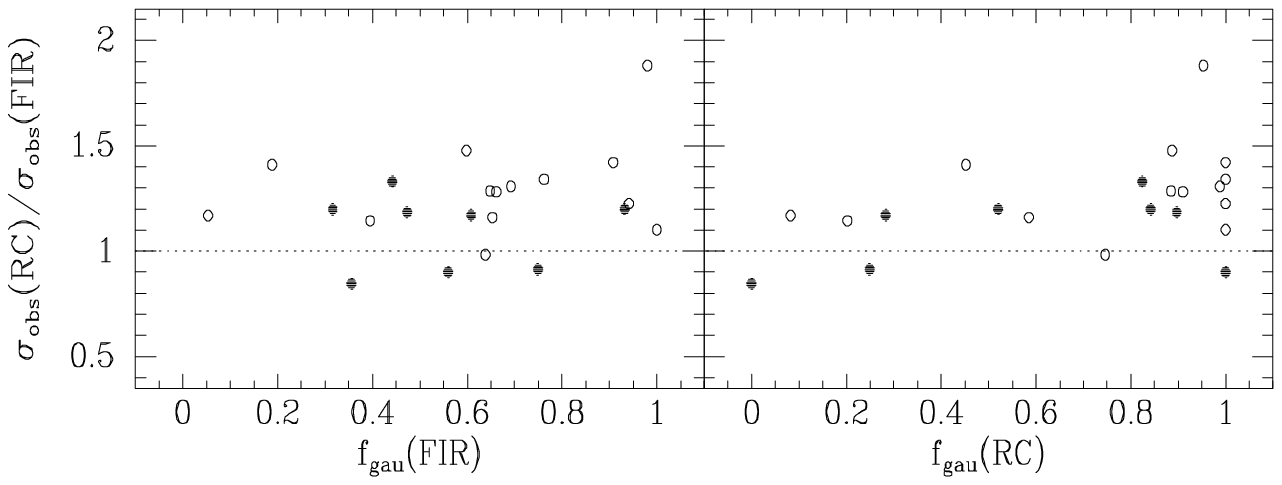}}
\vspace*{-1.5cm}
\caption{
The ratio of the scale lengths of RC and FIR emission is plotted against the
relative flux in the gaussian component (f$_{\rm gau}$). f$_{\rm gau}$ for
the FIR and RC emission is shown separately in the two panels. In a given
galaxy f$_{\rm gau}$ for FIR and RC deviates by more than 30\% in eight
galaxies which are identified by filled circles. Except in four galaxies
(NGC\,3628, 4192, 4321 and 4569),
the RC scale lengths are larger than the FIR scale lengths. 
}

\vspace*{-0.5cm}
\centerline{\psfig{figure=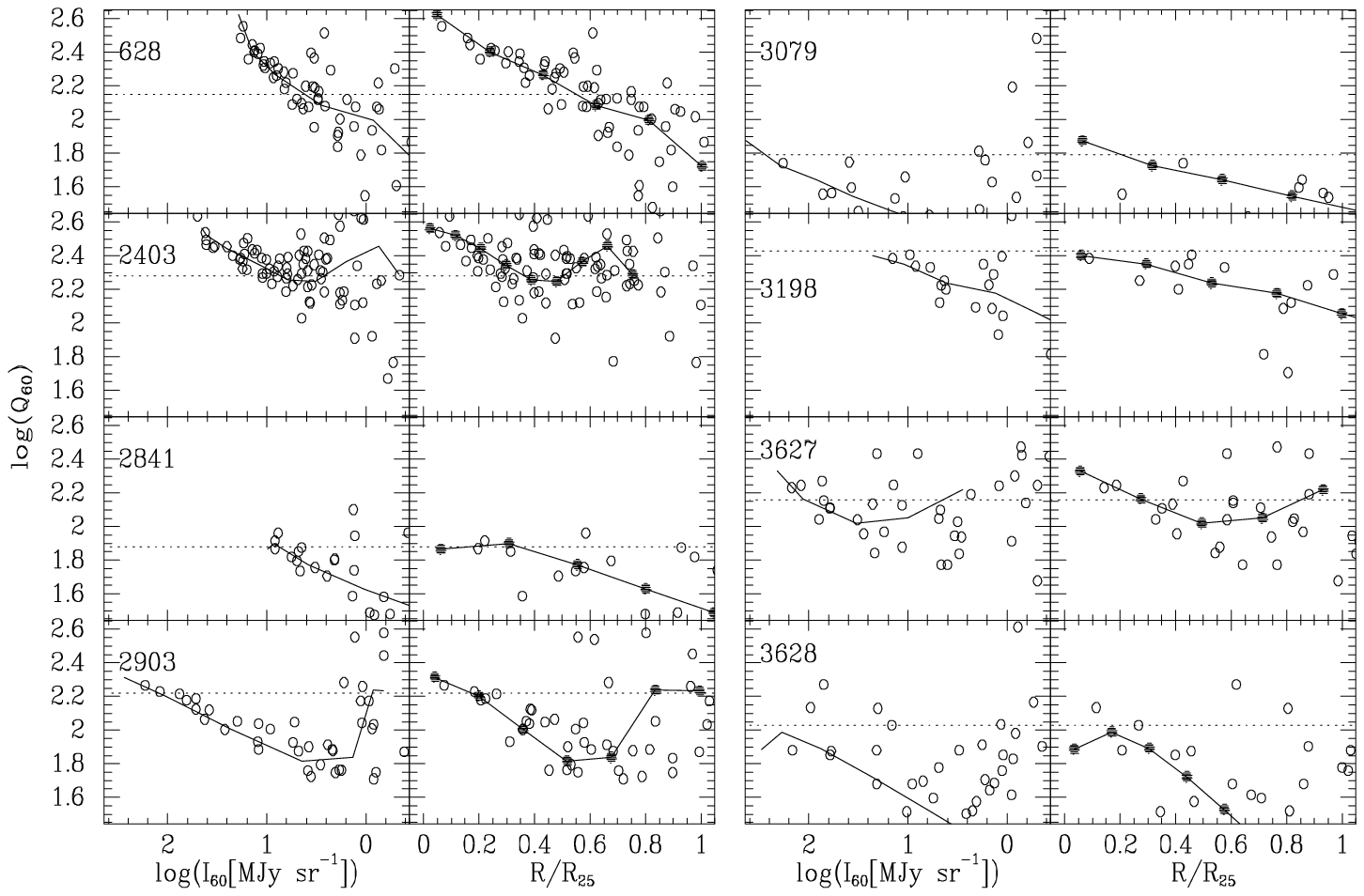,height=11.0cm}}
\caption{
\Qsixty\ values within individual galaxies are plotted against the 
60\micron\ intensity and the radial distance from the center. Open circles 
correspond to 
averages over 1\arcmin\ pixels in galaxies. Results from azimuthally
averaged radial profiles are shown by the thin line (left panel) and
thick lines (right panel). The radial profiles are sampled at 1\arcmin\ 
intervals, which are indicated by the solid dots on the right panel.
In general \Qsixty\ values show a tendency to decrease by about factors
2--3 within half the optical radius of the galaxy. 
}
\end{figure}

\begin{figure}[ht]
\vspace*{-2.5cm}
\centerline{\psfig{figure=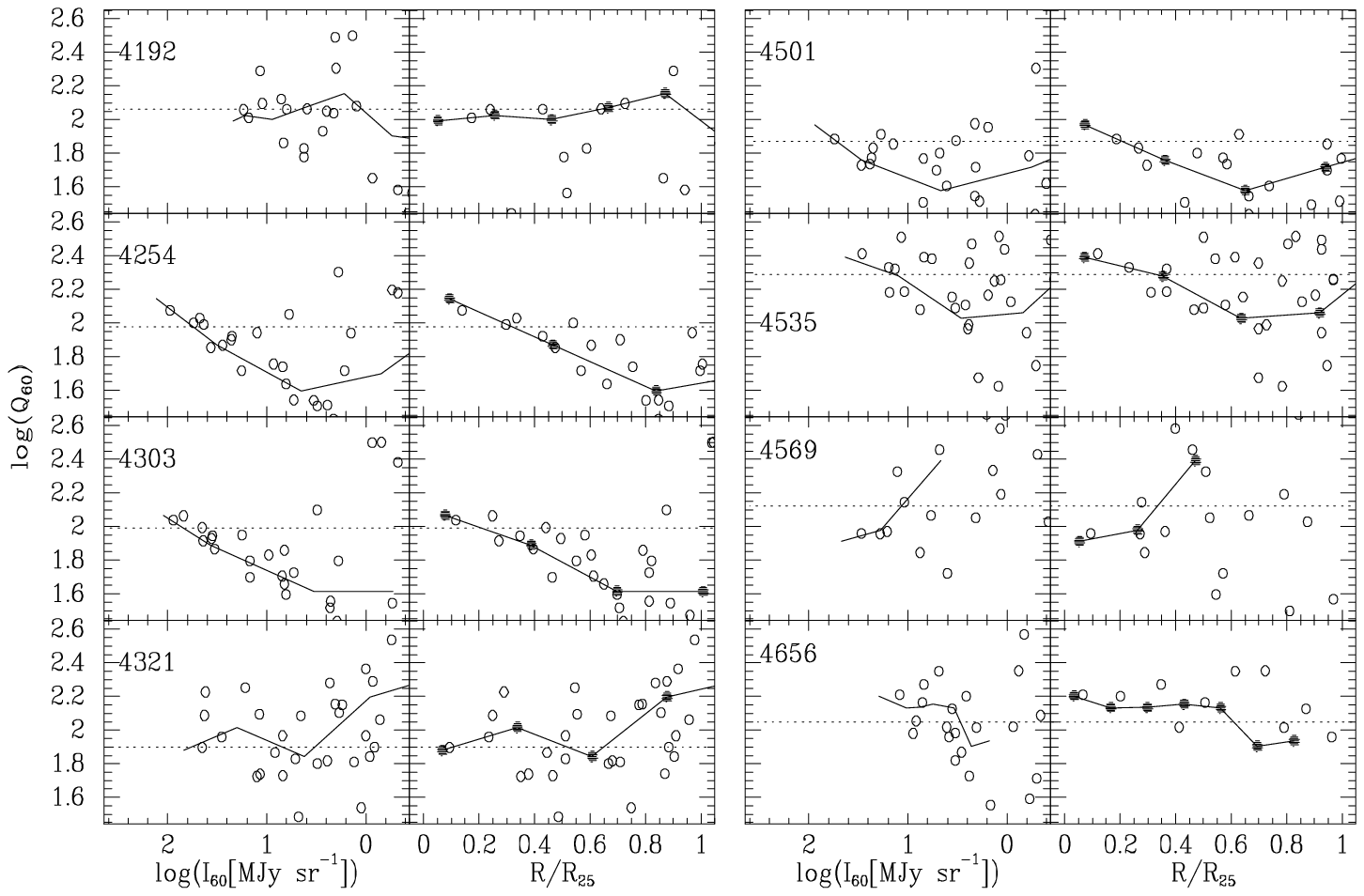,height=11.0cm}}
\vspace*{-2.0cm}
\centerline{\psfig{figure=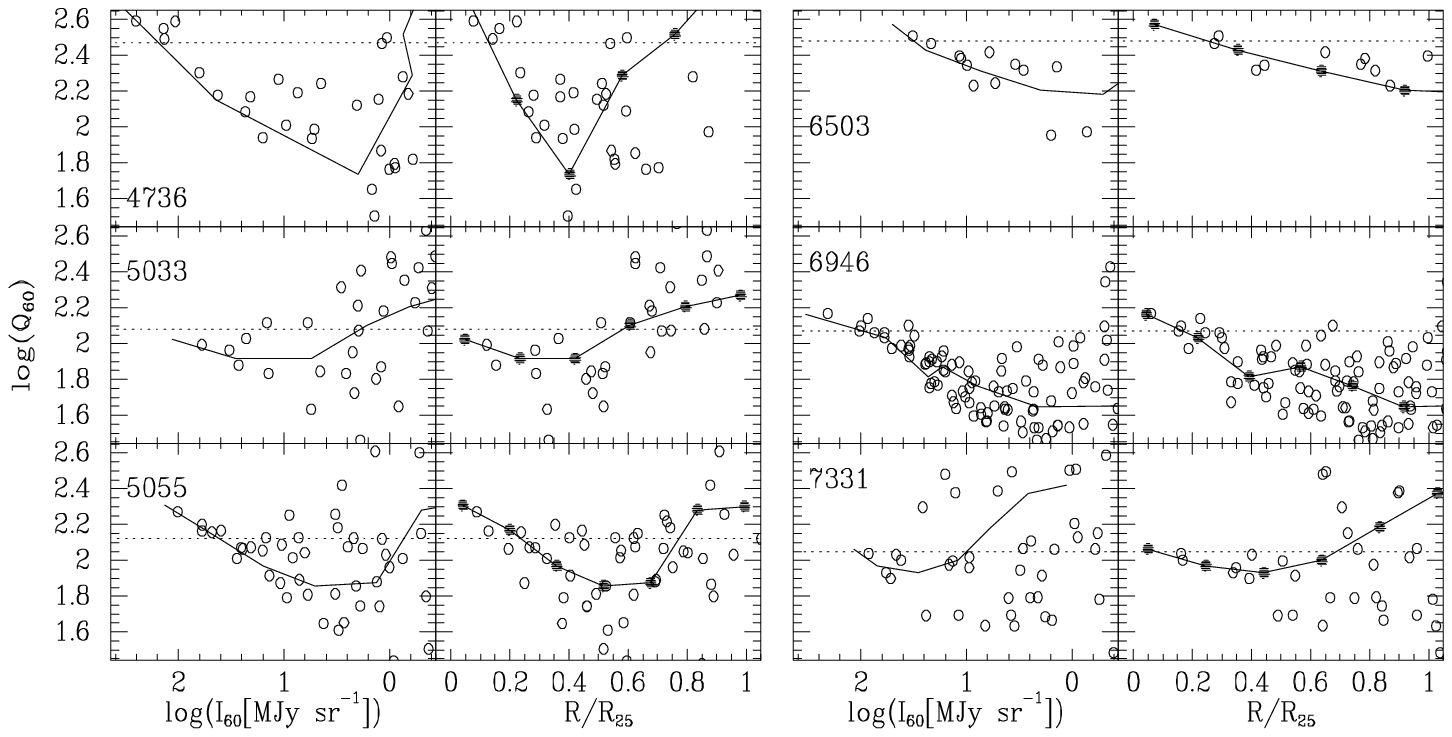,height=11.0cm}}
\vspace*{1.0cm}
\centerline{Fig.~8.--- Continued.}
\end{figure}

\twocolumn

\begin{figure}[ht]
\hspace*{1.0cm}
\centerline{\psfig{figure=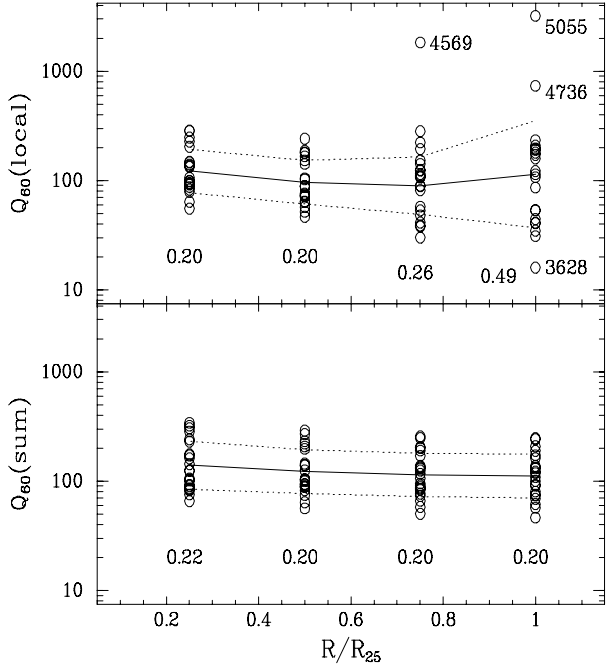,height=10.0cm}}
\caption{
\Qsixty\ values for the program galaxies are plotted at 25, 50, 75 and 
100\% of the disk radius (top panel). Mean values at each of these radii
are connected by a solid line. Dotted lines trace locations of 1-rms 
deviation from the mean. The values of the rms deviation (in log units)
at the four
radial distances as well as the NGC numbers of galaxies with maximum 
dispersion are noted. The bottom panel is similar to the top panel 
except that the plotted \Qsixty\ values are the result of integrating
the fluxes inside successively larger radial zones.
Hence the values at the
outer-most position correspond to the global \Qsixty\ and dispersion.
}
\end{figure}

\begin{figure}[ht]
\vspace*{-5.3cm}
\hspace*{0.5cm}
\centerline{\psfig{figure=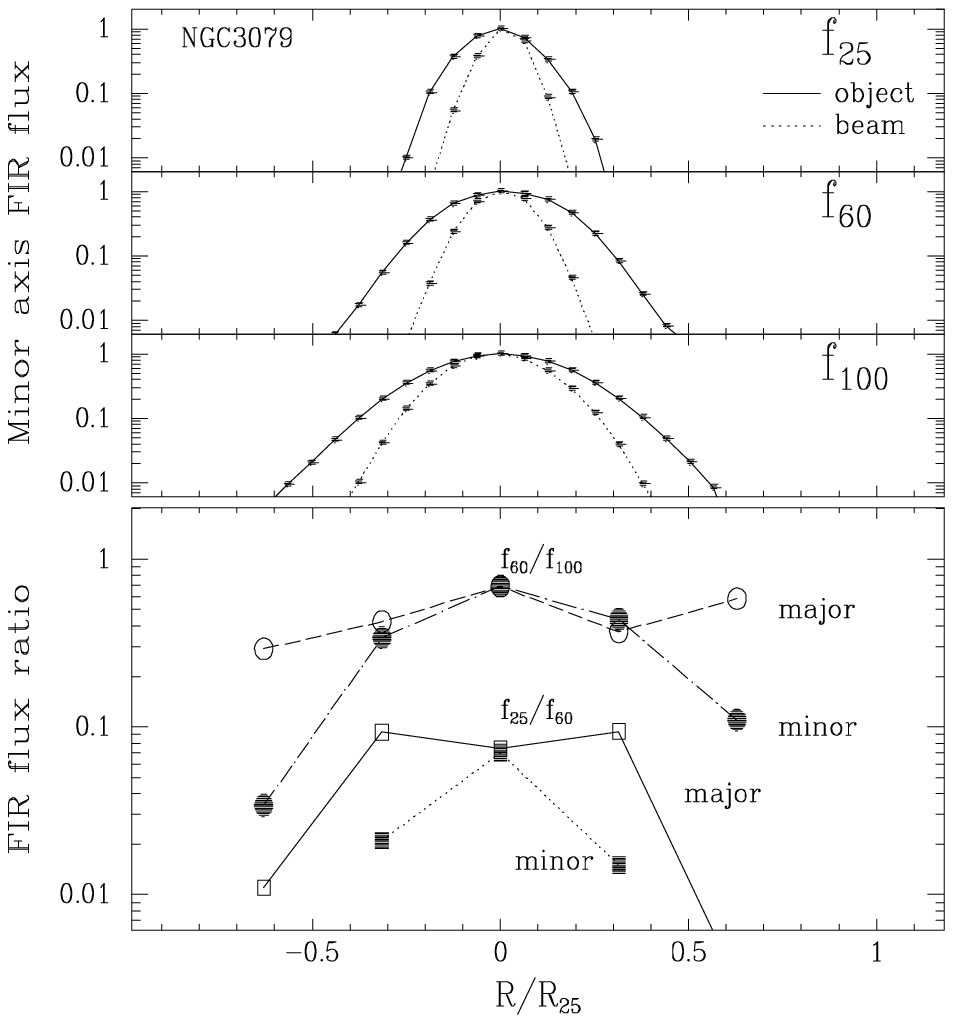,height=10.0cm}}
\caption{
The FIR flux ratios (which are indicators of dust temperature) in 25, 60 and 
100 \micron\ bands along the major and minor axes of the edge-on galaxy 
NGC\,3079 are plotted, in the bottom-most panel.  
The flux and beam profiles along the minor axis are plotted in the first
three panels. The negative x-axis corresponds
to the south-east side (see Fig.~1). Symbols denote the measured
values, which are averages over 1 arcmin pixels. The
connecting lines illustrate the steeper gradient along the minor
axis, especially for the hotter dust (f$_{25}$/f$_{60}$).
}
\end{figure}

\end{document}